\documentstyle[12pt]{article}

\def\o{\Omega}
\def\ph{\phi}

\def\pr{\partial }

\def\th{\theta}
\def\G{\Gamma}
\def\S{\Sigma}
\def\O{\Omega}
\def\Gd{G_{D+d}}
\def\Bd{B_{D+d}}
\def\tG{\tilde{G}_{D+d}}
\def\tB{\tilde{B}_{D+d}}
\def\td{\tilde{d}}
\newcommand{\be}{\begin{equation}} \newcommand{\ee}{\end{equation}}
\newcommand{\bea}{\begin{eqnarray}}\newcommand{\eea}{\end{eqnarray}}

\begin{document}
\baselineskip= 24 truept

\begin{titlepage}


\title{ Gauged String Actions and $O(\td, \td)$
Transformation}

\author{Alok Kumar$^1$\\
Theory Division, CERN, Geneva\\ and\\
Institute of Physics, Bhubaneswar-751005, INDIA$^2$}


\footnotetext[1]{e-mail:kumar@iopb.ernet.in}
\footnotetext[2]{Permanent address}

\date{}
\maketitle

\thispagestyle{empty}

\vskip 1in
\begin{abstract}

\vskip .3in

An $O(\td, \td)$ transformation is given which relates
the background configuration
of the ungauged string action to the gauged ones for a large class of
models discussed recently by Giveon and Rocek. Interestingly, the
transformation is background independent and has a unique
matrix representation in a given space-time dimension.

\end{abstract}

\leftline{CERN-TH.6530/92}
\leftline{IP/BBSR/92-44}
\leftline{June 1992}

\bigskip
\vfil

\end{titlepage}

\eject

Last year, in a series of papers, it was shown by Meissener and
Veneziano$[1]$, Sen and Hassan$[2,3]$ that the string
effective action in any
space-time dimension is invariant under an $O(\td, \td)$ group of
symmetry transformations when the background fields
are independent of $\td$ of the space-time coordinates.
Such transformations generate new classical solutions
from a given one, and have been conjectured to be a
generalization of the Narain construction$[4]$ to the curved
background. Indeed, they have been a useful tool for
generating several interesting nontrivial solutions, both the
cosmological$[5]$ as well as the static ones with curvature
singularity, such as rotating black holes$[6]$, charged branes$[2,3]$
etc.

On the other hand, the space-time geometry of the gauged WZW
models
has also been analyzed intesively since last year and gives an
exact conformal field therory representation of many interesting
target space structures such as two dimensional black holes$[7,8,
9,10]$,
three dimensional charged black string$[11, 12]$ and many other
solutions$[13]$. In this context, in order to formulate duality
in curved backgrounds $[14]$, a large class of
string actions and their abelian gauging was recently studied by
Giveon and Rocek$[15]$. The gauged string action of
ref.[15] corresponds
to the most general background configuration in a given space-time
dimension D which is independent of d of the coordinates. The
corresponding ungauged string action is the (D+d)-dimensional one with its
background parameterized by the elements of the backgrounds of the
gauged action. It has been pointed out in ref.[15] that the two
dimensional black hole as well as some other interesting classical
solutions belong to this class of backgrounds.
 We believe that, being a
large class, many more interesting classical solutions are likely
to fall into this category of models.

In this paper, we show an intimate connection between the two
approaches, namely the $O(\td, \td)$ transformation [1,2,3] and
the abelian gauging [15], used for
the construction of  string backgrounds. We give an
$O(\td, \td)$ transformation, with $\td = 2d$ in ref.[15],  which
relates the ungauged backgrounds to the gauged ones for the complete
class of models studied in ref.[15].  An interesting aspect
of the result is that the
corresponding $O(\td, \td)$ tansformation is represented by a
unique (D+d)-dimensional matrix for the whole class of models.
Under this transformation, the background configuration of the
ungauged string action transforms to a direct product of the one
corresponding to the gauged action and a set of free scalar fields
whose number is same as that of the gauged U(1) currents.
These transformations can therefore be interpreted as
a deformation from a conformal field theory $[\cal{F}]$, described
by the ungauged action of ref.[15], to another one $[{\cal{F}}/
{U(1)}^d] \times {U(1)}^d$.
The background independence of the transformation
may have relation with the   characteristic features
of the operator responsible for
such a deformation in the corresponding CFT.

We start by reviewing the relevant aspects of ref.[15]. It
was shown in ref.[15]
that the string action in a space-time dimension D,
with most general background independent
of d of the coordinates, can be constructed by gauging a class of
(D+d)-dimensional action with respect to its $U(1)_L^d \times
U(1)_R^d$ conserved currents.
  The ungauged action is written as
\be
	S_{D+d} = {1\over 2\pi}{\int d^2z \left[{(E_{D+{d}})}_{IJ}
\pr X^I \bar{\pr}X^J - {1\over 4}\ph_{D+d} R^{(2)} \right]}
\ee
where $X^I \equiv (x^a, \th_1^i, \th_2^i)$, a = 1...$(D-{d})$ and
i= 1...${d}$. The background $E_{IJ}$($x$) depends only on
coordinates $x^a$ and is written as,
\be
E_{D+{d}}\; = (\Gd - \Bd) =
	\;\left (\matrix {{\G} & {{\G}_1^T} & {0}\cr
{} & {} & {} \cr
{0} & {I_d} & {-I_d}\cr
{} & {} & {} \cr
{{\G}_2} & {2\S + I_d} & {I_d}\cr }\right ),
\ee
where $\G$, $\S$  and $\G_{1,2}$ are $(D-d)\times
(D-d)$, $d \times d$ and $d \times (D-d)$ dimensional
$x^a$ dependent matrices, respectively.
Subscript of the Identity matrix
in eqn.(2) and in the rest of the paper denotes
its dimension.
The (D+d)-dimensional background is, therefore, parameterized by
four matrices $\G$, $\G_{1,2}$, and $\S$.
The $\Gd$, $\Bd$ and $\phi_{D+d}$ are the background metric, antisymmetric
tensor and dilaton fields such that the action (1) is conformal invariant.
Therefore, these fields satisfy the equations
of motion of the string effective action in (D+d)-dimensions.

It has already been observed in ref.[15] that a constant shift in the
antisymmetric tensor in ($\th_1, \th_2$) space corresponds to the
addition of a total derivative in the action (1). We will later on
also show
that such shifts can be obtained by $O(\td, \td)$ transformation.
Taking these
facts into account, we ignore a shift of this type in the expression for
$E_{D+d}$ in eqn.(2) and write  $\Gd$ and $\Bd$ as,
\be
G_{D+{d}}\; =
	\;\left (\matrix {{\G_s} & {{1\over 2}{\G}_1^T} &
	{{1\over 2}{\G}_2^T}\cr
{} & {} & {} \cr
 {{1\over 2}{\G}_1} & {I_d} & {\S^T}\cr
{} & {} & {} \cr
{{1\over 2}{\G}_2} & {\S} & {I_d}\cr }\right ),
\ee
and
\be
B_{D+{d}}\; =
	\;\left (\matrix {{\G_a} & {-{1\over 2}{\G}_1^T} &
	{{1\over 2}{\G}_2^T}\cr
{} & {} & {} \cr
 {{1\over 2}{\G}_1} & {0} & {\S^T}\cr
{} & {} & {} \cr
 {-{1\over 2}{\G}_2} & {-\S} & {0}\cr }\right ),
\ee
where $\G_{s,a} = \pm {1\over 2} {(\G \pm \G^T)}$.
The action (1) is
invariant under
${U(1)}^d_L \times {U(1)}^d_R$ transformations.
By gauging these symmetries and
integrating out the gauge fields one obtains
\be
	S_{D} = {1\over 2\pi}{\int d^2z \left[{(E_{D})}_{AB}
\pr X^A \bar{\pr}X^B - {1\over 4}\ph_D R^{(2)} \right]}
\ee
where $E_D \equiv (G_D-B_D)$, $X^A \equiv (x^a, \th^i)$ and
$\th^i = (\th_1^i - \th_2^i)$.
 The resulting
D-dimensional background metric ($G_D$),
antisymmetric tensor ($B_D$)  and dilaton ($\ph_D$) are [15]:
\be
G_{D}\; =
	\;\left (\matrix { {[\G_s-{1\over 4}
	{{\G_1}^T}{(I_d+\S)}^{-1}{\G_2}} &
	{{1\over 2}[{{\G}_1^T}{(I_d+\S)}^{-1}} \cr
	{-{1\over 4}{{\G_2}^T}{(I_d+\S^T)}^{-1}{\G_1]}} &
	{-{{\G}_2^T}{(I_d+\S^T)}^{-1}]} \cr
	{} & {} \cr
 	{{1\over 2}[{(I_d+\S^T)}^{-1}{\G}_1} &
	{{1\over 2}{\left[{(I_d-\S)}{(I_d+\S)}^{-1} \right.}} \cr
	{-{(I_d+\S)}^{-1}{{\G}_2]}} &
	{\left. +{(I_d+\S^T)}^{-1}{(I_d-\S^T)}\right]}	\cr}
 \right ),
\ee
\be
B_{D}\; =
	\;\left (\matrix { {[\G_a+{1\over 4}{{\G_1}^T}{(I_d+\S)}^{-1}{\G_2}} &
	{-{1\over 2}{[{\G}_1^T}{(I_d+\S)}^{-1}} \cr
	{-{1\over 4}{{\G_2}^T}{(I_d+\S^T)}^{-1}{\G_1}]} &
	{+{{\G}_2^T}{(I_d+\S^T)}^{-1}]} \cr
	{} & {} \cr
 	{{1\over 2}[{(I_d+\S^T)}^{-1}{\G}_1} &
	{{1\over 2}{\left[-{(I_d-\S)}{(I_d+\S)}^{-1} \right.}} \cr
	{+{(I_d+\S)}^{-1}{{\G}_2}]} &
	{\left. +{(I_d+\S^T)}^{-1}{(I_d-\S^T)}\right]}	\cr}
 \right ),
\ee
and
\be
	\ph_D = \ph_{D+d} + log[det(I_d+\S)].
\ee
It has been pointed out[15] that, $G_D$, $B_D$ and $\ph_D$
obtained in this manner
are consistent D-dimensional string backgrounds.

We now come to the main point of the paper.
In this paper, we show that there is an $O(\td, \td)$
transformation [1,2,3], represented by the $2(D+{d})\times 2(D+{d})$
dimensional matrix:
\be
\o\;\equiv\;\left (\matrix { {I_{D-d}} & {0} & {0} & {0} &
	{0} & {0} \cr
{0} & {I_d\over 2} & {-{I_d\over 2}} & {0}
& {-{I_d\over 2}} & {-{I_d\over 2}}\cr
{0} & {I_d\over 2} & {{I_d\over 2}} & {0}
& {{I_d\over 2}} & {-{I_d\over 2}}\cr
{0} & {0} & {0} & {{I_{D-d}}}
& {0} & {0}\cr
{0} & {-{I_d\over 2}} & {-{I_d\over 2}} & {0}
& {{I_d\over 2}} & {-{I_d\over 2}}\cr
{0} & {I_d\over 2} & {-{I_d\over 2}} & {0}
& {{I_d\over 2}} & {{I_d\over 2}}\cr }\right )
\ee
which is the symmetry of the (D+d)-dimensional string efffective
action and transforms the
background fields, ($G_{D+d}$, $B_{D+d}$) to ($\tG$, $\tB$) respectively,
where
\be
	\tG\; =
	\;\left (\matrix {{G_D} &  {0}\cr
 	{0} & {I_d} \cr }\right ),
\ee
\be
	\tB\; =
	\;\left (\matrix {{B_D} &  {0}\cr
 	{0} & {0} \cr }\right ).
\ee
We also show that the dilaton $\ph_{D+d}$ transforms under the
above $O(\td, \td)$ transformation as
\be
	\ph_{D+d} \rightarrow \ph_D.
\ee
Therefore, as stated earlier, under the above $O(\td, \td)$ transformation
the backgrounds $(\Gd, \Bd, \ph_{D+d})$
for the ungauged string action transform to the gauged
one $( G_D, B_D, \ph_D)$ together with a set of free scalars.
It is remarkable to note that a background independent $\O$ in eqn.(9)
transforms the nontrivial expressions of eqns.(3-4) and (6-7) into
each other, especially since $O(\td, \td)$ does not act linearly
on the background graviton and antisymmetric tensor fields.

$\O$ in eqn.(9) satisfies the condition, ${\O}^T \eta \O = \eta$,
where
\be
	\eta\; = \;\left (\matrix{ {0} & {I_{D+d}} \cr
		{I_{D+d}} & {0} \cr }\right ).
\ee
Its action  on ($\Gd$, $\Bd$) is given by
a linear transformation on a matrix $M$:
\be
	\O M_{D+d} \O^T = \tilde{M}_{D+d},
\ee
where
\be
M_{D+{d}}\; =
	\;\left (\matrix {{G_{D+d}^{-1}} &
	 {-{G_{D+d}^{-1}}B_{D+d}}\cr
	{} & {} \cr
 	{B_{D+d}G_{D+d}^{-1}} &
	{ G_{D+d} - B_{D+d}G_{D+d}^{-1}B_{D+d}}\cr }\right ),
\ee
and  $\tilde{M}_{D+d}$ in eqn.(14) is a matrix of the same
form as $M_{D+d}$,
with ($G_{D+d}$, $\Bd$) replaced by ($\tG$, $\tB$).
The dilaton field $\ph_{D+d}$ transforms under this
transformation to [1,2,3]
\be
	\tilde{\ph}_{D+d} = \ph_{D+d} + {1\over 2}
		log\left({{det \Gd}\over {det G_D}}\right).
\ee
and it will be shown later on that $\tilde{\ph}_{D+d} = \ph_D$.

The proof that the transformation $\O$ in eqn.(9) is a symmetry of
the $(D+d)$- dimensional string effective action follows directly
from the results of ref.[3]. By setting the gauge field backgrounds
in ref.[3] to zero, $\O$ in eqn.(9) is easily seen to belong to the
set of symmetry transformations given in eq.(3.21) of ref.[3], when one
takes into account that $\eta$
in ref.[3] has a different form than ours.

Our major task now is to verify eqn.(14).
Algebraic complexity in this calculation arises mainly due to the
complicated forms of $\Gd^{-1}$ and $G_D^{-1}$ as well as other elements
of the matrices $M_{D+d}$, and $\tilde{M}_{D+d}$ in terms
of $\G$, $\G_1$, $\G_2$ and $\S$. For example,
\be
G_{D+\hat{d}}^{-1}\; =
	\;\left (\matrix {{x} & {y^T} & {z^T}\cr
 {y} & {A} & {B^T}\cr
 {z} & {B} & {C}\cr }\right ),
\ee
where
\bea
	x = &&x^T = {\left[ \G_s - {1\over 4} \G_2^T\G_2 -
	{1\over 4}\G_1^T{(I_d-\S^T\S)}^{-1}\G_1
	- {1\over 4}\G_2^T\S{(I_d-\S^T\S)}^{-1}\S^T\G_2 \right.}\cr
&& {\left.	 + {1\over 4}\G_1^T{(I_d-\S^T\S)}^{-1}\S^T\G_2
	+ {1\over 4}\G_2^T\S{(I_d-\S^T\S)}^{-1}\G_1 \right]}^{-1} \cr
	y =  && {1\over 2}{\left[-{(I_d-\S^T\S)}^{-1}\G_1 +
		{(I_d-\S^T\S)}^{-1}\S^T\G_2 \right]} x \cr
	z = &&  {1\over 2}{\left[\S{(I_d-\S^T\S)}^{-1}\G_1 -
		\G_2 -
		\S {(I_d-\S^T\S)}^{-1}\S^T\G_2 \right]} x \cr
	A = &&  {(I_d-\S^T\S)}^{-1} + y x^{-1} y^T \cr
	B = &&  {-\S {(I_d-\S^T\S)}^{-1} + z x^{-1} y^T} \cr
	C = &&  {I_d + \S {(I_d-\S^T\S)}^{-1} \S^T
		+ z x^{-1} z^T}
\eea

$\Gd^{-1}$ in eqn.(17) has been
obtained by using a general inversion
formula for a matrix of the type:
\be
  	D\; =
	\;\left (\matrix { {P} & {R^T} \cr
{R} & {Q} \cr }\right ),
\ee
then,
\be
  	D^{-1}\; =
	\;\left (\matrix { {{(P-R^T Q^{-1}R)}^{-1}} &
	{-{(P-R^T Q^{-1}R)}^{-1} R^T Q^{-1}} \cr
{} & {} \cr
{-Q^{-1} R {(P-R^T Q^{-1}R)}^{-1}} &
{Q^{-1} + Q^{-1} R {(P-R^T Q^{-1}R)}^{-1} R^T Q^{-1}} \cr }\right ).
\ee

Using the expressions for $\Gd$, $\Bd$ and $\Gd^{-1}$ in equations
(3), (4) and (17),
the  matrix $M_{D+d}$ in eqn.(15) can be written explicitly
in terms of matrices
$\G_{s,a}, \G_{1,2}, \S, x, y, z, A, B,$ and $C$.
The expression for $ (\Gd^{-1}\Bd)$ is given as,
\be
	-(\Gd^{-1}\Bd)\; =
	\;\left (\matrix {
{-x\G_a - {1\over 2}y^T\G_1 + {1\over 2}z^T\G_2} &
{{1\over 2}x\G_1^T + z^T\S} & {-{1\over 2}x\G_2^T - y^T\S^T}\cr
{} & {} & {} \cr
 {-y\G_a-{1\over 2}A\G_1 +{1\over 2}B^T\G_2} &
{{1\over 2}y\G_1^T + B^T\S} &
{-{1\over 2}y\G_2^T - A\S^T}\cr
{} & {} & {} \cr
 {-z\G_a-{1\over 2}B\G_1 +{1\over 2}C\G_2} &
{{1\over 2}z\G_1^T + C\S} &
{-{1\over 2}z\G_2^T - B\S^T}\cr }\right ),
\ee
An explicit expression for the matrix ($\Gd - \Bd\Gd^{-1}\Bd$) can
be obtained in a similar manner. Since it is
straightforward to obtain and not convenient to write in a compact
form, we do not give it here.

For further
calculations, we find it convenient to use a set of
identities which are  obtained from the requirements,
\be
\Gd\Gd^{-1} = \Gd^{-1}\Gd = I_{D+d},
\ee
where $\Gd$ and $\Gd^{-1}$ are given in eqns.(3) and (17) respectively.

Then, the left hand side of eqn.
(14) is computed by using the expression of $\O$ from eqn.(9)
and $M_{D+d}$. The calculation is
rather long and tedious. For simplifications we use the identities
obtained from eqn.(22). Finally,
for the left hand side of eqn.(14) one finds:
\be
\O{M_{D+{d}}}\O^T\; \equiv
	\;\left (\matrix {{{(\O M \O^T)}_{11}} & {{(\O M \O^T)}_{12}}\cr
{} & {} \cr
{[{{(\O M \O^T)}_{12}}]}^T & {{(\O M \O^T)}_{22}} \cr }\right ),
\ee
where
\be
{(\O M \O^T)}_{11}\; =
	\;\left (\matrix {{x} & {y^T-z^T} & {0} \cr
	{} & {} & {} \cr
 	{y-z} & {-I_d+ A + C} & {0} \cr
	{} & {-B-B^T} & {} \cr
	{} & {} & {} \cr
{0} & {0} & {I_d}\cr }\right ),
\ee
and
\be
{(\O M \O^T)}_{12}\;  =
	\;\left (\matrix {{[-x\G_a -{1\over 2}y^T\G_1 +
	{1\over 2}z^T\G_2]} &
{-(y^T+z^T)} & {0} \cr
{} & {} & {} \cr
 {[(-y+z)\G_a + {1\over 2}(- A\G_1  } &
{[-A-B^T} & {0} \cr
{+ B^T\G_2 + B\G_1 - C\G_2)]}
	& {+B+C]} & {} \cr
{} & {} & {} \cr
 {0} & {0} & {0}\cr }\right ),
\ee
The expression for ${(\O M \O^T)}_{22}$ is
again large and we do not give it here.
 One can then verify, using the expressions for
$\tG$ and $\tB$ in eqns.(10-11) and (6-7),
and the inversion formula (20), that,
\be
	{(\O M \O^T)}_{11} = \tG^{-1}.
\ee
In proving eqn.(26), we also used the following
simple identities,
\bea
{1\over 2}\left[(I_d - \S) {(I_d + \S)}^{-1} + \right.&&   \cr
\left.		{(I_d + \S^T)}^{-1}(I_d - \S^T) \right]
	&& = {(I_d+\S^T)}^{-1}
	(I_d - \S^T\S) {(I_d + \S)}^{-1},
\eea
and
\be
	-I_d + A + C -B - B^T =
		(I_d+\S){(I_d-\S^T\S)}^{-1}(I_d+\S^T)
		+ (y-z) x^{-1} (y^T - z^T).
\ee
Similarly one gets,
\bea
	{(\O M \O^T)}_{12} = && -\tG^{-1}\tB \cr
	{(\O M \O^T)}_{22} = && \tG - \tB \tG^{-1} \tB .
\eea
After verifying eqn.(14),
we now evaluate $\tilde{\ph}_{D+d}$ in eqn.(16).
We note that the determinant of the matrix $D$ in
eqn.(19) is given as
\be
Det(D) =  Det (Q). Det (P- R^T Q^{-1} R)
\ee
Using this expression one gets,
\be
	Det(\Gd) = {{Det(I_d-\S^T\S)}\over  Det(x)}
\ee
and
\be
	Det(\tG) = {{Det [ {(I_d+\S^T)}^{-1} (I_d-\S^T\S) {(I_d+\S)}^{-1}]}
		\over Det(x)}.
\ee

Then, for $\tilde{\ph}_{D+d}$, we have
\bea
	\tilde{\ph}_{D+d} = &&\ph_{D+d} + {1\over 2}
		log\left({{det \Gd}\over {det G_D}}\right) \cr
		  = && \ph_{D+d} + log [det(I_d+\S)] \cr
		  = && \ph_D,
\eea
which is the desired result in eqn.(12).

Finally, we like to mention that a constant shift in the antisymmetic
tensor $\Bd$ in ($\th_1, \th_2$) space can be obtained by an
$O(\td, \td)$ transformation,
\be
  	\O_s\; =
	\;\left (\matrix { {I_{D-d}} & {} & {} & {} \cr
{} & {I_{2d}} & {} & {} \cr
{} & {} & {I_{D-d}} & {} \cr
{} & {b} & {} & {I_{2d}} \cr }\right ),
\ee
where b is the constant shift in the antisymmetric tensor in
$(\th_1, \th_2)$ space. This justifies such shift in the
expression for $\Bd$ in eqn.(4).

To conclude, we have established that, for the
whole class of models in ref.[15],
background field configuration for the  unaguged
string action is related to
the gauged ones by a background independent
 $O(\td, \td)$ transformation. A simple
application of the result is that the $SL(2, R)$ WZW model can be
transformed by an $O(2, 2)$ transformation of the above type
to the uncharged
black string[11]. We expect that these results being general, can
be applied in various other circumstances.
It will be interesting to investigate whether the
results of this paper can be generalized even further, such as to
heterotic strings, nonabelian gaugings etc.

\vskip 1cm

\noindent{\bf Acknowledgement:} I would like to thank  M. Porrati and
S. Wadia for discussions and comments and the theory
division at CERN for its hospitality.

\vfil
\eject

\baselineskip 12pt
\vfil
\eject


\begin{thebibliography}{30.}
\bibitem[1]{} K. A. Meissner and G. Veneziano, Phys. Lett. {\bf B267},
33 (1991); Mod. Phys. Lett. {\bf A6}, 3397 (1991).

\bibitem[2]{} A. Sen, Phys. Lett. {\bf B271}, 295 (1991);
	{\bf B274}, 34 (1991).

\bibitem[3]{} S. Hassan and A. Sen, Nucl. Phys. {\bf B375},
	103, (1992).

\bibitem[4]{} K. Narain, Phys. Lett. {\bf B169}, 41 (1985);
	K. Narain, M. Sarmadi and E. Witten, Nucl. Phys.
	{\bf B279}, 369 (1987).

\bibitem[5]{} M. Gasperini, J. Maharana and G. Veneziano,
	Phys. Lett. {\bf B272}, 277 (1991); M. Gasperini and
	G. Veneziano, CERN Preprint CERN-TH-6321/91.

\bibitem[6]{} J. Horne and G. Horowitz,
	Santa Barbara Preprint UCSBTH-92-11;
	A. Sen, Tata Institute Preprint, TIFR/TH/92-20.


\bibitem[7]{} E. Witten, Phys. Rev. {\bf D44}, 314 (1991)

\bibitem[8]{} G. Mandal, A. Sengupta and S. Wadia, Mod. Phys. Lett.
	{\bf A6}, 1685 (1991).

\bibitem[9]{} R. Dijgraaf, E. Verlinde and H. Verlinde,
Institute for Advanced Study preprint IASSNS-HEP-91/22 (1991);
I. Bars, Univ. of Southern California, Los
Angeles preprint USC-91-HEP-B3 (1991).

\bibitem[10]{} N. Ishibashi, M. Li and A. R. Steif, Phys. Rev. Lett.
{\bf 67}, 3336 (1991);
S. Khastgir and A. Kumar, Mod. Phys. Lett. {\bf
A6}, 3365 (1991); S. Kar, S. Khastgir and A. Kumar,
Bhubaneswar
Preprint IP/BBSR/91-51 (to appear in Mod. Phys.
Lett. A); M. McGuian, C. Nappi and S. Yost, Preprint
IASSNS-HEP-91/57.

\bibitem[11]{} J. Horne and G. Horowitz, Nucl. Phys. {\bf B
360}, 197 (1992).

\bibitem[12]{} S. Kar and A. Kumar, Bhubaneswar Preprint,
	IP/BBSR/92-18.

\bibitem[13]{} I.Bars and K. Sfetsos, Univ. of Southern California, Los
Angeles preprint USC-91-HEP-B6 (1991);
P. Horava, Phys. Lett. {\bf B278} 101 (1992); D. Lust
and C. Kounnas, CERN Preprint CERN-TH-6494/92; S. Mahapatra, Tata
Institute Preprint, TIFR/TH/92-28.

\bibitem[14]{} E. B. Kiritsis,  Mod. Phys. Lett. {\bf A6}, 2871 (1991);
M. Rocek and E. Verlinde, Institute for Advanced Study preprint
IASSNS-HEP-91/68 (1991); P. Ginsparg and F. Quevedo, Los Alamos
Preprint, LA-UR-92-640.


\bibitem[15]{} A. Giveon and M. Rocek, Institute for Advanced Study
Preprint, IASSNS-91/84.




\end{thebibliography}
\end{document}